\documentstyle[12pt]{ioplppt} 

\eqnobysec

\begin{document} 

\title{Finite-size corrections of an integrable chain with alternating spins}[
Finite-size corrections]
\author{ B - D D\"orfel\footnote{E-mail:
doerfel@qft2.physik.hu-berlin.de} and St Mei\ss ner\footnote{E-mail:
meissner@qft2.physik.hu-berlin.de}} \address{Institut f\"ur Physik,
Humboldt-Universit\"at , Theorie der Elementarteilchen\\
Invalidenstra\ss e 110, 10115 Berlin, Germany}

\begin{abstract} 
In this paper we calculate the finite-size corrections of an anisotropic 
integrable spin chain, consisting of spins $s=1$ and $s=\frac{1}{2}$. The 
calculations are done in two regions of the phase diagram with respect to the 
two couplings $\bar{c}$ and $\tilde{c}$. In case of conformal invariance we
obtain the final answer for the ground state and its lowest excitations, which
generalizes earlier results.
\end{abstract} 


\maketitle 

\section{Introduction} 
In 1992 de Vega and Woynarovich constructed the first example of a spin chain 
with alternating spins of the values $s=\frac{1}{2}$ and $s=1$ \cite{devega} 
on the basis of the well-known XXZ($\frac{1}{2}$) model. We call this model
XXZ($\frac{1}{2},1$). 
Later on a lot of interesting generalizations have been presented 
\cite{aladim1,aladim2,martins}. After de Vega {\it et al} 
\cite{devega1,devega2} we have studied the XXZ($\frac{1}{2},1$) model in two 
subsequent publications \cite{meissner,doerfel}. In our last paper 
\cite{doerfel} we determined the ground state for different values of the two 
couplings $\bar{c}$ and $\tilde{c}$ (for the details see section 3 of that 
paper). Disregarding two singular lines we have found four regions in the 
($\bar{c},\tilde{c}$)-plane which can be divided into two classes. The division
is made with respect to the occurance of finite Fermi zones for Bethe ansatz 
roots. The two regions with infinite fermi zones only are well studied 
\cite{devega,meissner} in the framework of Bethe ansatz. On that basis we 
consider the finite-size corrections for the ground state and its lowest 
excitations using standard techniques \cite{eckle,woy,hamer}. It is remarkable 
that they allow to obtain an explicit answer only in the conformally invariant 
cases, which are contained in the two regions considered. The results can 
easily be compared with the predictions of conformal invariance.

The paper is organized as follows.
Definitions are reviewed in section 2. In section 3 we 
calculate the finite-size corrections for both couplings negative. The same is 
done in section 4 for positive couplings. Here it was necessary to set 
$\bar{c}=\tilde{c}$ to obtain explicit answers. Section 5 contains the 
interpretation of the results and our conclusions.

\section{Description of the model}
We consider the Hamiltonian of a spin chain of length $2N$ with $N$ even
\begin{equation}\label{ham}
{\cal H}(\gamma) = \bar{c} \bar{\cal H}(\gamma) + \tilde{c} \tilde{\cal H}
(\gamma).
\end{equation}
The two Hamiltonians can (implicitly) be found in paper \cite{devega}, they
both contain a two-site and a three-site coupling part. Their explicit 
expressions are rather lengthy and do not provide any further insights. They
include a XXZ-type anisotropy parametrized by $e^{i \gamma}$, we restrict 
ourselves to $0<\gamma<\pi/2$. The isotropic limit XXX($\frac{1}{2},1$) is 
contained in \cite{aladim1}. The two real coupling constants $\bar{c}$ and 
$\tilde{c}$ dominate the qualitative behaviour of the model. The interaction 
favours antiparallel orientation of spins, for equal signs of the couplings 
its character resembles ordinary $XXZ$ model. A new kind of competition comes 
in for different signs of couplings where the ground state is still singlet 
but with a much more involved structure.

The Bethe ansatz equations (BAE) determining the solution of the model are
\begin{equation}\label{bae}
\fl \left( \frac{\sinh(\lambda_j+i\frac{\gamma}{2})}{\sinh(\lambda_j-i
\frac{\gamma}{2})}
\frac{\sinh(\lambda_j+i\gamma)}{\sinh(\lambda_j-i\gamma)} \right)^N =
-\prod_{k=1}^{M}\frac{\sinh(\lambda_j-\lambda_k+i\gamma)}{\sinh(\lambda_j-
\lambda_k-i\gamma)},\qquad j=1\dots M.
\end{equation}
One can express energy, momentum and spin projection in terms of BAE roots 
$\lambda_j$:
\begin{eqnarray}\label{en}
E = \bar{c} \bar{E} + \tilde{c} \tilde{E},
\nonumber\\
\bar{E} = - \sum_{j=1}^{M} \frac{2\sin\gamma}
{\cosh2\lambda_j - \cos\gamma},
\nonumber\\
\tilde{E} = - \sum_{j=1}^{M} 
\frac{2\sin2\gamma}{\cosh2\lambda_j - \cos2\gamma},
\end{eqnarray}
\begin{equation}\label{mom}
P =\frac{i}{2}\sum_{j=1}^{M} \left\{ \ln \left(\frac{\sinh(\lambda_j+i\frac{
\gamma}{2})}{\sinh(\lambda_j-i\frac{\gamma}{2})} \right) + 
\ln \left( \frac{\sinh(\lambda_j+i\gamma)}{\sinh(\lambda_j-i\gamma)} \right) 
\right\},
\end{equation}
\begin{equation}\label{spin}
S_z = \frac{3N}{2} - M.
\end{equation}
We have defined energy and momentum to vanish for the ferromagnetic state.
The momentum operator was chosen to be half of the logarithm of the 2-site 
shift operator \cite{aladim1} which is consistent with taking the length of the
system as $2N$ instead of $N$.

\section{Calculation of finite-size corrections for negative couplings}
In section 3 of paper \cite{doerfel} we have carried out a detailed analysis of
the thermodynamic Bethe ansatz equations (TBAE) at zero temperature and 
obtained the ground state.

We found a large antiferromagnetic region in the ($\bar{c},\tilde{c}$)-plane 
(depending on $\gamma$) where the ground state is formed by roots with 
imaginary parts $\frac{\pi}{2}$, the so-called ($1,-$) strings. The Fourier
transform of their density is given by \cite{meissner}
\begin{equation}\label{(1,-)dens}
\hat{\rho}_0(p)=\frac{1+2\cosh(p\gamma/2)}{2\cosh(p(\pi-\gamma)/2)}.
\end{equation} 
Depending on the signs of $\tilde{c}$ and $\bar{c}$ the region is described by
the connection of three parts:
\begin{enumerate}
\item[a)] 
\begin{eqnarray*}
\tilde{c}\leq0,\bar{c}\leq0;
\end{eqnarray*}
\item[b)] 
\begin{eqnarray*}
\tilde{c}<0,\bar{c}>0\\
\frac{\bar{c}}{|\tilde{c}|}\leq\frac{1}{2\cos\tilde{\gamma}} \qquad \mbox{for} 
\qquad 0<\gamma\leq\frac{2\pi}{5}\\
\frac{\bar{c}}{|\tilde{c}|}\leq2\cos\tilde{\gamma} \qquad \mbox{for} \qquad 
\frac{2\pi}{5}\leq\gamma<\frac{\pi}{2};
\end{eqnarray*}
\item[c)]
\begin{eqnarray}\label{phase}
\tilde{c}>0,\bar{c}<0\nonumber\\
\frac{|\bar{c}|}{\tilde{c}}\geq\frac{8\cos^3\tilde{\gamma}}{4\cos^2\tilde{
\gamma}-1} \qquad \mbox{for} \qquad 0<\gamma\leq\frac{\pi}{3}\nonumber\\
\frac{|\bar{c}|}{\tilde{c}}\geq\frac{2}{\cos\tilde{\gamma}} \qquad \mbox{for} 
\qquad \frac{\pi}{3}\leq\gamma<\frac{\pi}{2}.
\end{eqnarray}
\end{enumerate}
Here for shortness we have introduced
\begin{equation}
\tilde{\gamma}=\frac{\pi\gamma}{2(\pi-\gamma)}.
\end{equation}
We shall now calculate the finite-size corrections for the ground state and its
excitations. In paper \cite{meissner} the structure of excitations in the 
framework of BAE roots was obtained for $\tilde{c}<0,\bar{c}<0$. Our results 
immediately apply to the whole region (\ref{phase}), because we had to ensure 
only that the ground state consists of ($1,-$) strings which follows from TBAE.

Because we are interested in the lowest excitations only, we disregard the 
bound states \cite{meissner} and consider those excitations given by holes in 
the ground-state distribution which are located right (or left) from the real 
parts of all roots. The number of those holes we call $H^+$ ($H^-$). We follow
the standard techniques developed in \cite{eckle} and \cite{woy}.

For transparency we employ the notations of \cite{hamer} as much as possible.
We decompose 
\begin{equation}
\sigma_N=\rho_0^{(1)} + \rho_0^{(2)} + \Delta\sigma_N
\end{equation}
where the upper index describes the two terms on the RHS of formula 
(\ref{(1,-)dens}). The basic equations are then
\begin{eqnarray}\label{DE}
\fl
\frac{\Delta E_N}{2N}\equiv e_N = \bar{c} \pi \int_{-\infty}^{\infty} d\lambda
\rho_0^{(1)}(\lambda) \left\{ \frac{1}{N} \sum_k \delta(\lambda-\lambda_k) 
- \sigma_N(\lambda) \right\}\nonumber\\
+ \tilde{c} \pi \int_{-\infty}^{\infty} d\lambda
\rho_0^{(2)}(\lambda) \left\{ \frac{1}{N} \sum_k \delta(\lambda-\lambda_k) 
- \sigma_N(\lambda) \right\}
\end{eqnarray}
and
\begin{equation}\label{Dsigma}
\Delta\sigma_N (\lambda) = - \int_{-\infty}^{\infty} d\mu \bar{p}(\lambda-\mu)
\left\{ \frac{1}{N} \sum_k \delta(\lambda-\lambda_k) - \sigma_N(\lambda) 
\right\}
\end{equation}
where the Fourier transform of the kernel $\bar{p}(\lambda)$ is given by
\begin{equation}\label{kernel}
\bar{P}(\omega) = - \frac{\sinh((\pi/2-\gamma)\omega)}{2\sinh(\omega\gamma/2)
\cosh(\omega(\pi/2-\gamma/2))}.
\end{equation}
The summation on the RHS of (\ref{DE}) and (\ref{Dsigma}) is carried out over 
the real parts of all the roots (without the holes). Using Euler-MacLaurin 
formula equations (\ref{DE}) and (\ref{Dsigma}) are rewritten as usual, e.g.
\begin{eqnarray}\label{Dsigma_rew}
\fl
\sigma_N(\lambda)-\rho_0^{(1)}(\lambda)-\rho_0^{(2)}(\lambda) =
\nonumber\\
\int_{\Lambda^+}^{\infty}d\mu\sigma_N(\mu)
\bar{p}(\lambda-\mu) - \frac{1}{2N}\bar{p}(\lambda-\Lambda^+) + \frac{1}{12N^2}
\frac{1}{\sigma_N(\Lambda^+)}\bar{p}(\lambda-\Lambda^+)
\nonumber\\
\left( + \int^{\Lambda^-}_{-\infty}d\mu\sigma_N(\mu)
\bar{p}(\lambda-\mu) - \frac{1}{2N}\bar{p}(\lambda-\Lambda^-) - \frac{1}{12N^2}
\frac{1}{\sigma_N(\Lambda^-)}\bar{p}(\lambda-\Lambda^-) \right)\nonumber\\
\end{eqnarray}
Here $\Lambda^+$ ($\Lambda^-$) is the real part of the largest (smallest) root.
For $\lambda\geq\Lambda^+$ the part in round brackets can be omitted and 
equation (\ref{Dsigma_rew}) after a shift converts into a standard Wiener-Hopf
problem to be solved. In the expression for $\Delta E_N$ both parts have to be 
kept, so we need the solution for $\lambda\geq\Lambda^+$ and $\lambda\leq
\Lambda^-$, which are simply related (but not equal) by symmetry.

For the solution with $\lambda\geq\Lambda^+$ we define as usual
\begin{equation}\label{Xpm}
X_{\pm}(\omega)=\int_{-\infty}^{\infty} e^{i\omega\lambda}\sigma_N^{\pm}
(\lambda+\Lambda^+) d\lambda,
\end{equation}
\begin{eqnarray}\label{spm}
\sigma_N^{\pm}(\lambda+\Lambda^+)=\left\{
\begin{array}{ll}
\sigma_N(\lambda+\Lambda^+) & \mbox{for}\quad\lambda \quad{> \atop <} \quad 0\\
0			    & \mbox{for}\quad\lambda \quad{< \atop >} \quad 0\\
\end{array}
\right. .
\end{eqnarray}
After Fourier transformation equation (\ref{Dsigma_rew}) takes the form
\begin{eqnarray}\label{e_ft}
X_-(\omega) + (1-\bar{P}(\omega))(X_+(\omega)-\bar{C}(\omega))=
\bar{F}_+(\omega)+\bar{F}_-(\omega)-\bar{C}(\omega)
\end{eqnarray}
where we have marked all given functions of our problem by a bar. 
$\bar{F}_{\pm}(\omega)$ are defined as above using instead of $\sigma_N$ the
sum $\rho_0^{(1)}+\rho_0^{(2)}$. Further
\begin{equation}\label{bar_C}
\bar{C}(\omega)=\frac{1}{2N} + \frac{i\omega}{12 N^2 \sigma_N(\Lambda^+)}.
\end{equation}
Now we have to factorize the kernel 
\begin{equation}\label{k}
[1-\bar{P}(\omega)] = \bar{G}_+(\omega)\bar{G}_-(\omega)
\end{equation}
with $\bar{G}_{\pm}(\omega)$ holomorphic and continuous in the upper and lower
half-plane respectively. Noticing that 
\begin{equation}\label{new_kernel}
\bar{P}(\omega,\gamma)=K(\omega,\pi-\gamma)
\end{equation}
where $K$ is the analogous function in paper \cite{hamer} we take the 
factorization from there.
\begin{eqnarray}\label{fac1}
\fl
\bar{G}_+(\omega)=\sqrt{2\gamma}\Gamma\left(1-\frac{i\omega}{2}\right)
e^{\bar{\psi}(\omega)}\left[ \Gamma\left(\frac{1}{2}-\frac{i(\pi-\gamma)\omega}
{2\pi}  \right) \Gamma\left(   \frac{1}{2}-\frac{i\gamma\omega}
{2\pi}\right) \right]^{-1} = \bar{G}_-(-\omega),
\end{eqnarray}
\begin{eqnarray}\label{fac2}
\bar{\psi}(\omega)=\frac{i\omega}{2}\left[\ln\left(\frac{\pi}{\gamma}\right)
- \frac{\pi-\gamma}{\pi} \ln\left(\frac{\pi-\gamma}{\gamma}\right)\right].
\end{eqnarray}
It is chosen to fulfill 
\begin{eqnarray}\label{asy}
\bar{G}_+(\omega)\stackrel{|\omega|\to\infty}{\sim}1+\frac{\bar{g}_1}{\omega}
+\frac{\bar{g}_1^2}{2\omega^2} + {\cal O}\left(\frac{1}{\omega^3}\right)
\end{eqnarray}
where 
\begin{equation}\label{g1}
\bar{g}_1=\frac{i}{12}\left(2+\frac{\pi}{\pi-\gamma}-\frac{2\pi}{\gamma}\right)
\end{equation}
After the necessary decomposition 
\begin{equation}\label{dec}
\bar{G}_-(\omega)\bar{F}_+(\omega) = \bar{Q}_+(\omega)+\bar{Q}_-(\omega)
\end{equation}
equation (\ref{e_ft}) has the desired form
\begin{eqnarray}\label{form}
\fl
\frac{X_+(\omega)-\bar{C}(\omega)}{\bar{G}_+(\omega)}-\bar{Q}_+(\omega)=
\bar{Q}_-(\omega)-\bar{G}_-(\omega)\left[X_-(\omega)+\bar{C}(\omega)
-\bar{F}_-(\omega)\right] \equiv  \bar{P}(\omega)
\end{eqnarray}
leading to an entire function $\bar{P}(\omega)$ given by its asymptotics.
\begin{eqnarray}\label{bar_p}
\bar{P}(\omega)=\frac{i\bar{g}_1}{12 N^2 \sigma_N(\Lambda^+)} - \frac{1}{2N}
-\frac{i\omega}{12 N^2 \sigma_N(\Lambda^+)}.
\end{eqnarray}
Equation (\ref{form}) yields the solution for $X_+(\omega)$:
\begin{eqnarray}\label{sol}
X_+(\omega)=\bar{C}(\omega)+\bar{G}_+(\omega)\left[\bar{P}(\omega)+\bar{Q}_+
(\omega)\right].
\end{eqnarray}
for our purposes it is sufficient to put 
\begin{eqnarray}\label{f+}
\bar{F}_+(\omega)=\frac{e^{\pi\Lambda^+/(\pi-\gamma)}}{\pi-i\omega(\pi-\gamma)}
(1+2\cos\tilde{\gamma})
\end{eqnarray}
and hence
\begin{equation}\label{bar_q+}
\bar{Q}_+(\omega)=\frac{\bar{G}_+(i\pi/(\pi-\gamma))e^{-\pi\Lambda^+/
(\pi-\gamma)}(1+2\cos\tilde{\gamma})}{\pi-i\omega(\pi-\gamma)}.
\end{equation}
Next we must determine by normalization the value of the integral 
\begin{eqnarray*}
\int_{\Lambda^+}^{\infty}\sigma_N(\lambda)d\lambda=z_N(\infty)-z_N(\Lambda^+).
\end{eqnarray*}
After a thorough analysis we found for our case that the relation
\begin{equation}\label{holes}
\frac{H}{2}=\frac{H^++H^-}{2}=\left[ \nu S + \frac{1}{2} \right]
\end{equation}
holds where $\nu=\frac{\gamma}{\pi}$. Nevertheless we shall not claim formula 
(\ref{holes}) for all possible states \cite{woy}, especially we expect effects 
like those described in \cite{juettner} for higher excitations. Then we have
\begin{eqnarray}\label{z}
z_N(\pm\infty)-z_N(\Lambda^{\pm})&=&\pm\frac{1}{N}\left( \frac{1}{2}+\nu S_z 
\pm \frac{H^+-H^-}{2} \right)\nonumber\\
&\equiv& \pm\frac{1}{N}\left( \frac{1}{2} \pm \Delta^{\pm} \right)
\end{eqnarray}
yielding the important equation
\begin{equation}\label{imp1}
\fl
\frac{\bar{G}_+(i\pi/(\pi-\gamma))e^{-\pi\Lambda^+/(\pi-\gamma)}
(1+2\cos\tilde{\gamma})}{\pi} = \frac{1}{2N} - \frac{i\bar{g}_1}{12N^2\sigma_N
(\Lambda^+)} + \frac{1}{N} \frac{1}{\sqrt{2\nu}}\Delta^+.
\end{equation}
The other normalization equation is obviously
\begin{equation}\label{imp2}
\fl
\sigma_N(\Lambda^+)=\frac{\bar{g}_1^2}{24N^2\sigma_N(\Lambda^+)} + 
\frac{i\bar{g}_1}{2N} + \frac{\bar{G}_+(i\pi/(\pi-\gamma))}{\pi-\gamma}
e^{-\pi\Lambda^+/(\pi-\gamma)}(1+2\cos\tilde{\gamma}).
\end{equation}
Now we can proceed in the usual way keeping in mind the changes arising 
especially from the last two equations.
\begin{eqnarray}\label{DE_1}
\fl
\frac{\Delta E_N}{2N}= -\frac{\pi}{\pi-\gamma}(\bar{c}+2\tilde{c}\cos
\tilde{\gamma}) \bar{G}_+\left(\frac{i\pi}{\pi-\gamma}\right)
\left[ \bar{P}\left(\frac{i\pi}{\pi-\gamma}\right) + \bar{Q}_+
\left(\frac{i\pi}{\pi-\gamma}\right) \right] e^{-\pi\Lambda^+/(\pi-\gamma)}
\nonumber\\
+ \left(  \Lambda^+\leftrightarrow\Lambda^-\right).
\end{eqnarray}
After some algebra using equations (\ref{imp1}) and (\ref{imp2}) this turns out
as
\begin{eqnarray}\label{DE_2}
\fl
\frac{\Delta E_N}{2N}= -\frac{\pi^2}{\pi-\gamma}\frac{(\bar{c}+2\tilde{c}\cos
\tilde{\gamma})}{1+2\cos\tilde{\gamma}}\times\nonumber\\
\times 
\left\{ \left( -\frac{1}{24N^2} + \frac{1}{4N^2\nu}(\Delta^+)^2 \right) 
+ \left( -\frac{1}{24N^2} + \frac{1}{4N^2\nu}(\Delta^-)^2 \right)\right\}.
\end{eqnarray}
Finally, for further interpretation we put it in the form
\begin{eqnarray}\label{DE_3}
\fl
\frac{\Delta E_N}{2N}= -\frac{2\bar{c}+4\tilde{c}\cos\tilde{\gamma}}
{1+2\cos\tilde{\gamma}}\frac{\pi}{\pi-\gamma}
\left\{ -\frac{\pi}{6}\frac{1}{4N^2} + \frac{2\pi}{4N^2}\left(\frac{S_z^2\nu}
{2}+\frac{\Delta^2}{2\nu} \right) \right\}.
\end{eqnarray}
with $\Delta=(H^+-H^-)/2$ as an integer number.

The momentum correction $\Delta P_N$ is obtained from relation (\ref{DE}) after
substituting the hole energy 
$\varepsilon_h=-2\bar{c}\pi\rho_0^{(1)}-2\tilde{c}\pi\rho_0^{(2)}$ by the hole 
momentum $p_h(\lambda)=\frac{1}{2}\arctan(\sinh(\pi\lambda)/(\pi-\gamma))+
\arctan(\sinh(\pi\lambda)/(\pi-\gamma))/\cos\tilde{\gamma})+$ const (see 
\cite{doerfel}).
 
Comparing the asymptotics for large $\lambda$ of both $\varepsilon_h(\lambda)$
and $p_h(\lambda)$ gives the speed of sound and helps to shorten the 
calculation of $\Delta P_N$. Therefore 
\begin{equation}\label{DP}
\frac{\Delta P_N}{2 N} = \frac{\pi}{2}\left\{ \frac{1}{4 N^2 \nu} 
\left[ (\Delta^-)^2 - (\Delta^+)^2 \right] \right\} + \mbox{const}.
\end{equation}
We are not interested in the constant term, being some multiple of $\pi$.

Finally
\begin{equation}\label{DP_res}
\Delta P_N = -\frac{2\pi}{2N} S_z \Delta.
\end{equation}

The interpretation of our result will be given in section 5. We stress once 
more, that to obtain equations (\ref{DE_3}) and (\ref{DP_res}) it was not 
necessary to put $\bar{c}=\tilde{c}$. The coupling constants are only constrained
to stay in region (\ref{phase}).

\section{Calculation of finite-size corrections for positive couplings}
Now we consider region $\bar{c}>0,\tilde{c}>0$ and rely on the analysis of
paper \cite{devega}. 

The ground state is given by two densities, $\sigma_N^{(1/2)}(\lambda)$ for the
real roots and $\sigma_N^{(1)}(\lambda)$ for the real parts of the ($2,+$) 
strings. One has 
\begin{equation}\label{dens}
\sigma_{\infty}^{(1/2)}(\lambda) = \sigma_{\infty}^{(1)}(\lambda) = 
\frac{1}{2\gamma\cosh(\pi\lambda/\gamma)} \equiv s(\lambda).
\end{equation} 
The physical excitations are holes in those distributions. As in section 3 we
consider only holes situated right (or left) from all roots. With the usual
technique and the results of \cite{devega} we have obtained after some lengthy
but straightforward calculations the basic system for the density corrections
\begin{eqnarray}\label{dens_corr}
\Delta\sigma_N^{(1/2)}(\lambda) = 
&-&\int_{-\infty}^{\infty}d\mu\:s(\lambda-\mu)
\left\{\frac{1}{N}\sum_{j=1}^{M_1}\delta(\mu-\xi_j)-\sigma_N^{(1)}(\mu)\right\}
\nonumber\\
\Delta\sigma_N^{(1)}(\lambda) = 
&-&\int_{-\infty}^{\infty}d\mu\:s(\lambda-\mu)\left\{ \frac{1}{N}\sum_{i=1}^
{M_{1/2}} \delta(\mu-\lambda_i) - \sigma_N^{(1/2)}(\mu) \right\}
\nonumber\\
&-&\int_{-\infty}^{\infty}d\mu\:r(\lambda-\mu)
\left\{\frac{1}{N}\sum_{j=1}^{M_1}\delta(\mu-\xi_j)-\sigma_N^{(1)}(\mu)\right\}
.
\end{eqnarray}
We have denoted the real roots by $\lambda_i$ (their number is $M_{1/2}$) and
the real parts of the strings by $\xi_j$ (number $M_1$). The function 
$r(\lambda)$ is given via its Fourier transform
\begin{equation}\label{q}
R(\omega)=\frac{\sinh(\omega(\pi-3\gamma)/2)}{2\sinh(\omega(\pi-2\gamma)/2)
\cosh(\omega\gamma/2)}.
\end{equation}
The energy correction takes the form
\begin{eqnarray}\label{en_corr}
\fl
\frac{\Delta E_N}{2N} = 
&-&\pi\bar{c} \int_{-\infty}^{\infty}d\lambda s(\lambda) \left\{ \frac{1}{N}
\sum_{i=1}^{M_{1/2}}\delta(\lambda-\lambda_i)-\sigma_N^{(1/2)}(\lambda)\right\}
\nonumber\\
\fl
&-&\pi\tilde{c} \int_{-\infty}^{\infty}d\lambda s(\lambda)\left\{\frac{1}{N}
\sum_{j=1}^{M_1}\delta(\lambda-\xi_j) - \sigma_N^{(1)}(\lambda) \right\}.
\end{eqnarray}
Once more we shall follow \cite{hamer}. The maximal (minimal) real roots we 
call $\Lambda^{\pm}_{1/2}$ and for the strings we use $\Lambda^{\pm}_{1}$ 
respectively. Instead of one $C(\omega)$ we have now $C_1(\omega)$ and 
$C_{1/2}(\omega)$ generalized in an obvious way. The same applies to $F(\omega)$.
The main mathematical problem is the factorization of a matrix kernel
\begin{eqnarray}\label{fact_mat}
\left( 1-K(\omega) \right)^{-1} = G_+(\omega) G_-(\omega) \qquad \mbox{with}
\nonumber\\
G_-(\omega)=G_+(-\omega)^T
\end{eqnarray}
(see \cite{devega1}) and
\begin{eqnarray}\label{k_mat}
K(\omega)=\left(  
\begin{array}{cc}
0 & S(\omega) e^{-i\omega\left(\Lambda_1^+-\Lambda_{1/2}^+\right)} \\
S(\omega) e^{i\omega\left(\Lambda_1^+-\Lambda_{1/2}^+\right)} & R(\omega)
\end{array}
\right).
\end{eqnarray}
$G_+$ is now a matrix function and $G_+^T$ stands for its transposition. The 
two component vector $Q_+(\omega)$ is (see equation (\ref{dec}))
\begin{eqnarray}\label{q+_vec}
Q_+(\omega)=\frac{G_+(i\pi/\gamma)^T}{\pi-i\omega\gamma} \left(
\begin{array}{c}
e^{-\pi\Lambda_{1/2}^+/\gamma} \\
e^{-\pi\Lambda_{1}^+/\gamma}
\end{array}
\right).
\end{eqnarray}
As usual we define the constant matrices $G_1$ and $G_2$ by
\begin{equation}\label{const_mat}
G_+(\omega) \stackrel{|\omega|\to\infty}{\longrightarrow} 
1 + G_1 \frac{1}{\omega} + G_2 \frac{1}{\omega^2} 
+ {\cal O}\left(\frac{1}{\omega^3}\right)
\end{equation}
and as before one has 
\begin{equation}
G_2=\frac{1}{2}G_1^2.
\end{equation}
The two component vector $P(\omega)$ is then
\begin{eqnarray}\label{p_vec}
P(\omega) = 
\left(
\begin{array}{c}
-\frac{1}{2N}-\frac{i\omega}{12N^2\sigma_N^{(1/2)}(\Lambda_{1/2}^+)}\\
-\frac{1}{2N} - \frac{i\omega}{12N^2\sigma_N^{(1)}(\Lambda_1^+)}
\end{array}
\right)
+ G_1
\left(
\begin{array}{c}
\frac{i}{12N^2\sigma_N^{(1/2)}(\Lambda_{1/2}^+)} \\
\frac{i}{12N^2\sigma_N^{(1)}(\Lambda_1^+)}
\end{array}
\right)
\end{eqnarray}
and therefore the shifted densities are expressed in the form
\begin{eqnarray}\label{f_dens}
\left(
\begin{array}{c}
X^+_{1/2}(\omega) \\
X^+_{1}(\omega)
\end{array}
\right) =
\left(
\begin{array}{c}
C_{1/2}(\omega) \\
C_{1}(\omega)
\end{array}
\right) + G_+(\omega)\left[ P(\omega)+Q_+(\omega) \right].
\end{eqnarray}
Now it is necessary to find the analogue of equation (\ref{z}) for the two
counting functions. Here it would be necessary to consider different cases 
depending on the fractions of $\nu S_z / N$ or $2\nu S_z / N$. From our 
experience we know that the result of the finite-size corrections does not 
depend on those fractions, while relations like (\ref{holes}) obviously do.
Being interested only in the former we shall proceed as easy as possible and
consider only the case with vanishing fractions.
\begin{equation}\label{z1}
z_N^{(1/2)}(\pm\infty)-z_N^{(1/2)}(\Lambda_{1/2}^{\pm})=
\pm\frac{1}{N}\left( \frac{1}{2} - \nu S_z + H_{1/2}^{\pm} \right),
\end{equation}
\begin{equation}\label{z2}
z_N^{(1)}(\pm\infty)-z_N^{(1)}(\Lambda_{1}^{\pm})=
\pm\frac{1}{N}\left( \frac{1}{2} - 2\nu S_z + H_{1}^{\pm} \right).
\end{equation}
Easy counting leads to expressions for the numbers of the holes
\begin{eqnarray}\label{h_num}
H_{1}=2 S_z
\nonumber\\
H_{1/2}=2 S_z + 2 M_1 -N.
\end{eqnarray}
We expect their modifications for non-vanishing fractions. We stress that both 
numbers are even.

Equation (\ref{imp1}) is now more complicated
\begin{eqnarray}\label{imp1_new}
\fl
\frac{G_+(i\pi/\gamma)^T}{\pi} \left(
\begin{array}{c}
e^{-\pi\Lambda_{1/2}^+/\gamma} \\
e^{-\pi\Lambda_{1}^+/\gamma}
\end{array}
\right) = G_+^{-1}(0)B^+ +
\left(
\begin{array}{c}
\frac{1}{2N} \\
\frac{1}{2N}
\end{array}
\right) - i G_1
\left(
\begin{array}{c}
\frac{1}{12N^2\sigma_N^{(1/2)}(\Lambda_{1/2}^+)} \\
\frac{1}{12N^2\sigma_N^{(1)}(\Lambda_1^+)}
\end{array}
\right)
\end{eqnarray}
with the definitions
\begin{eqnarray}\label{b+-}
\fl
B^{\pm} =
\left(
\begin{array}{c}
B_1^{\pm} \\
B_2^{\pm}
\end{array}
\right) = \frac{1}{N}
\left(
\begin{array}{c}
-\nu S_z + H_{1/2}^{\pm} \\
-2\nu S_z + H_1^{\pm}
\end{array}
\right) =
\frac{1}{N}
\left(
\begin{array}{c}
S_z -\nu S_z + M_1 - \frac{N}{2} \pm \Delta^{(1/2)} \\
S_z -2\nu S_z \pm \Delta^{(1)}
\end{array}
\right) 
\end{eqnarray}
and 
\begin{equation}\label{delta}
\Delta^{(i)} = \frac{H_i^+-H_i^-}{2}.
\end{equation}
The other normalization condition is obviously
\begin{eqnarray}\label{imp2_new}
\fl
\left(
\begin{array}{c}
\sigma_N^{(1/2)}(\Lambda^+_{1/2}) \\
\sigma_N^{(1)}(\Lambda^+_{1})
\end{array}
\right) = \frac{G_1^2}{2}
\left(
\begin{array}{c}
\frac{1}{12N^2\sigma_N^{(1/2)}(\Lambda_{1/2}^+)} \\
\frac{1}{12N^2\sigma_N^{(1)}(\Lambda_1^+)}
\end{array}
\right) + iG_1
\left(
\begin{array}{c}
\frac{1}{2N} \\
\frac{1}{2N}
\end{array}
\right) + \frac{G_+(i\pi/\gamma)^T}{\gamma}
\left(
\begin{array}{c}
e^{-\pi\Lambda_{1/2}^+/\gamma} \\
e^{-\pi\Lambda_{1}^+/\gamma}
\end{array}
\right).\nonumber\\
\end{eqnarray}
After combining equations (\ref{imp2_new}) and (\ref{imp1_new}) we obtain from 
equation (\ref{en_corr})
\begin{eqnarray}\label{en_corr_new}
\fl
\frac{\Delta E_N}{2N} = \frac{\pi}{\gamma}
\left(
\begin{array}{c}
\bar{c} e^{\pi\Lambda_{1/2}^+/\gamma} \\
\tilde{c} e^{\pi\Lambda_{1}^+/\gamma}
\end{array}
\right)^T G_+\left(\frac{i\pi}{\gamma}\right) \left[ - \frac{1}{2}
\left(
\begin{array}{c}
\frac{1}{2N} \\
\frac{1}{2N}
\end{array}
\right) + \left( \frac{1}{2} i G_1 + \frac{\pi}{\gamma} \right)
\left(
\begin{array}{c}
\frac{1}{12N^2\sigma_N^{(1/2)}(\Lambda_{1/2}^+)} \\
\frac{1}{12N^2\sigma_N^{(1)}(\Lambda_1^+)}
\end{array}
\right) \right.
\nonumber\\
\left.
+ \frac{1}{2} G_+^{-1}(0) B^+ \right]
+ (\Lambda^+\leftrightarrow\Lambda^-,B^+ \leftrightarrow B^-).
\end{eqnarray}
This result is valid for any positive $\bar{c}$ and $\tilde{c}$. As in paper 
\cite{devega1} no further progress can be made unless the factorization is
explicitly known which is not the case up to now. For $\bar{c}=\tilde{c}=c$ 
(conformal invariance) the problem simplifies and the final answer can be 
obtained.

After some lengthy calculations one arrives at
\begin{eqnarray}\label{final1}
\fl
\frac{\Delta E_N}{2N} = \frac{\pi^2 c}{\gamma} \left[ -\frac{1}{12N^2} 
+ \frac{1}{2} \left(  \left(  B_1^+ - B_2^+ \right)^2 + B_1^+B_2^+ - 
(B_2^+)^2 \frac{\pi-3\gamma}{2(\pi-2\gamma)} \right) \right] + 
(B^+ \leftrightarrow B^-)\nonumber\\
\end{eqnarray}
which can be brought into the form
\begin{eqnarray}\label{final2}
\fl
\frac{\Delta E_N}{2N} = \frac{2\pi c}{\gamma} \left\{ -\frac{\pi}{6} 
\frac{2}{4N^2} + \frac{2\pi}{4N^2} \left[ \frac{1}{4}(1-2\nu)S_z^2 
+ \frac{1}{4} \left(  H_{1/2} - \frac{H_1}{2}  \right)^2 
\right. \right.
\nonumber\\
\left. \left.
+(\Delta^{(1/2)})^2 - \Delta^{(1/2)}\Delta^{(1)} + \frac{1}{2} 
(\Delta^{(1)})^2 \frac{1-\nu}{1-2\nu} \right] \right\}.
\end{eqnarray}
For $\bar{c}\not=\tilde{c}$ we expect the result to be much more complicated 
but of the same order $1/N^2$.

As above the momentum correction is given from relation (\ref{en_corr}) after
replacing $\varepsilon_h^{(1/2)}=2\bar{c}\pi s(\lambda)$ by 
$p_h^{(1/2)}=\arctan e^{\pi\lambda/\gamma}$ and 
$\varepsilon_h^{(1)}=2\tilde{c}\pi s(\lambda)$ by 
$p_h^{(1)}=\arctan e^{\pi\lambda/\gamma}$. The values of the hole momenta are 
taken from \cite{devega} where an additionel factor $\frac{1}{2}$ must be 
introduced to take into account our definition of momenta (\ref{mom}).

As in section 3 comparing the asymptotics for $\bar{c}=\tilde{c}$ gives the 
speed of sound and together with equation (\ref{final1}) the momentum 
correction
\begin{eqnarray}\label{mom_corr}
\fl
\frac{\Delta P_N}{2N} = \frac{\pi}{2} \left\{ \frac{1}{2}\left[  
\left(  B_1^- - B_2^- \right)^2 - \left(  B_1^+ - B_2^+ \right)^2 
\right. \right. 
\nonumber\\
\left. \left. + B_1^-B_2^-
- B_1^+B_2^+ - ( (B_1^-)^2-(B_1^+)^2 ) \frac{\pi-3\gamma}{2(\pi-2\gamma)} 
\right] \right\} + \mbox{const}.
\end{eqnarray}
Disregarding the constant (multiple of $\pi$) we have
\begin{eqnarray}\label{mom_corr2}
\Delta P_N = \frac{\pi}{2} \left\{ -\Delta^{(1/2)} \left( H_{1/2}-\frac{H_1}{2}
 \right) -\Delta^{(1)}\left( \frac{H_1}{2}-\frac{H_{1/2}}{2}\right) \right\}.
\end{eqnarray}

\section{Conclusions}
In sections 3 and 4 we have determined the finite-size corrections of our model
for two different cases. Equations (\ref{DE_3}) and (\ref{DP_res}) give the 
result for region (\ref{phase}), while equations (\ref{final2}) and 
(\ref{mom_corr2}) are valid for $\bar{c}=\tilde{c}=c>0$. In both cases we have 
conformal invariance. That seems to be the reason why during the calculations a 
lot of terms cancel each other and the result is considerably simple.

From the asymptotics of $\varepsilon_h$ and $p_h$ we find
\begin{equation}\label{sos1}
v_s = - \frac{2\pi}{\pi-\gamma} \frac{\bar{c}+2\tilde{c}\cos\tilde{\gamma}}
{1+2\cos\tilde{\gamma}} > 0
\end{equation}
for the speed of sound and from relation (\ref{DE_3}) the value of $1$ for the
central charge.
Formula (\ref{sos1}) generalizes the former result obtained for equal (but 
negative) couplings \cite{doerfel}. Analogously follows
\begin{equation}\label{sos2}
v_s = \frac{2c\pi}{\gamma}
\end{equation}
and the central charge equals $2$ \cite{devega}.

For completeness we mention the heat capacities per site at low temperature
given by
\begin{equation}\label{hc1}
C = \frac{c_v T \pi}{3 v_s}
\end{equation}
where we have denoted the central charge by $c_v$ to avoid confusions. 
Therefore
\begin{equation}\label{hc2}
C = - \frac{1+2\cos\tilde{\gamma}}{\bar{c}+2\tilde{c}\cos\tilde{\gamma}}
\frac{(\pi-\gamma)T}{6}
\end{equation}
and
\begin{equation}\label{hc3}
C = \frac{\gamma T}{3 c}
\end{equation}
in agreement with former results \cite{doerfel}. Formula (\ref{hc2}) 
generalizes our calculations for $\bar{c}=\tilde{c}$.

The dimensions $x_n$ and the spins $s_n$ of the primary operators follow from
formulae (\ref{DE_3}) and (\ref{DP_res}).
\begin{equation}\label{po1}
x_n = \frac{S_z^2 \nu}{2} + \frac{\Delta^2}{2\nu}
\end{equation}
\begin{equation}\label{sop1}
s_n = S_z |\Delta|
\end{equation}
for negative coupling. It is interesting to compare this with the result for 
the XXZ($\frac{1}{2}$) model, where $\nu$ is simply replaced by $1-\nu$. 
Formulae (\ref{po1}) and (\ref{sop1}) have to be understood in the way that for 
general excited states (arbitrary holes and complex roots) $S_z$ and $\Delta$ are
replaced by more complicated integer numbers \cite{woy}.

For positive couplings
\begin{eqnarray}\label{po2}
\fl
x_n = \frac{1}{4}(1-2\nu)S_z^2 + \frac{1}{4} \left( H_{1/2}-\frac{H_1}{2} 
\right)^2 + \left( \Delta^{(1/2)} \right)^2 - \Delta^{(1/2)}\Delta^{(1)}
+ \frac{1}{2} \left( \Delta^{(1)} \right)^2 \frac{1-\nu}{1-2\nu}.
\end{eqnarray}
When $\nu\to0$ the first two terms agree with paper \cite{martins} while the 
other terms are connected with the asymmetry of the state which was not 
considered there. The dimension of a general primary operator depends on four
integer numbers. The second of them measures an asymmetry between the number of
holes among real roots or strings respectively. Once more, for more complicated
states the integers in equation (\ref{po2}) are replaced by other ones 
depending on the concrete structure of the state. We mention that relation 
(\ref{po2}) can be "diagonalized" to resemble the expression of two models both
of central charge $1$.
\begin{eqnarray}\label{decomp}
\fl
x_n = \frac{1}{2} \frac{(1-2\nu)}{2} S_z^2 + \frac{1}{2} 
\frac{\left( H_{1/2}-H_1/2 \right)^2}{2} + \frac{1}{2} 2\Delta_1^2
+ \frac{1}{2} \frac{2}{1-2\nu} \Delta_2^2 
\end{eqnarray}
with
\begin{eqnarray}\label{exp}
\Delta_1 = \Delta^{(1/2)} - \frac{\Delta^{(1)}}{2},
\nonumber\\
\Delta_2 = \frac{\Delta^{(1)}}{2}.
\end{eqnarray}
Expression (\ref{decomp}) becomes even more symmetric if one remembers the 
first equation of (\ref{h_num}) and the definition (\ref{delta}). Then twice a
certain number of holes is linked with its appropriate asymmetry.

From formula (\ref{mom_corr2}) the spins of the primary operators are 
\begin{equation}\label{sop2}
s_n = \left| \Delta^{(1/2)} \left( H_{1/2} - \frac{H_1}{2} \right)
+ \Delta^{(1)} \left( \frac{H_1}{2} - \frac{H_{1/2}}{2} \right) \right|
\end{equation}
and after using (\ref{exp}) 
\begin{equation}\label{sop3}
s_n = \left| \Delta_1 \left( H_{1/2} - \frac{H_1}{2} \right)
+ \Delta_{2} S_z \right|
\end{equation}
with the same symmetry as relation (\ref{decomp}).

Finally, we shall determine the magnetic suszeptibilities per site at zero 
temperature and vanishing field from our finite-size results. That can be done,
because the states we have considered include those with minimal energy for a 
given $S_z$ (magnetization) \cite{yangyang}. Differentiating twice the energy 
with respect to $S_z$ gives the inverse susceptibility.

Hence
\begin{equation}\label{sus1}
\chi = \frac{\pi-\gamma}{4\pi\gamma} \left( -\frac{1+2\cos\tilde{\gamma}}
{\bar{c}+2\tilde{c}\cos\tilde{\gamma}} \right) = \frac{1}{v_s}\frac{1}{2\gamma}
\end{equation}
and
\begin{equation}\label{sus2}
\chi = \frac{1}{2c\pi} \frac{1}{\pi-2\gamma}=\frac{1}{v_s}\frac{1}{\pi-2\gamma}
\end{equation}
for the two cases respectively in agreement with earlier results 
\cite{martins,devega1,doerfel}. Expression (\ref{sus1}) for general couplings 
in region (\ref{phase}) had not been derived before.

\section*{Acknowledgment}
One of us (St M) would like to thank H. J. de Vega for helpful discussions.


\section*{References}
\begin{thebibliography}{99}
\bibitem{devega}de Vega H J and Woynarovich F 1992
                 {\it J. Phys. A: Math. Gen.} {\bf 25} 4499
\bibitem{aladim1}Aladim S R and Martins M J 1993
		 {\it J. Phys. A: Math. Gen.} {\bf 26} L529
\bibitem{aladim2}Aladim S R and Martins M J 1993
		 {\it J. Phys. A: Math. Gen.} {\bf 26} 7301
\bibitem{martins}Martins M J 1993
		 {\it J. Phys. A: Math. Gen.} {\bf 26} 7287
\bibitem{devega1}de Vega H J, Mezincescu L and Nepomechie R I 1994
                 {\it Phys. Rev} {\bf B49} 13223
\bibitem{devega2}de Vega H J, Mezincescu L and Nepomechie R I 1994
                 {\it Int. J. Mod. Phys.} {\bf B8} 3473
\bibitem{meissner}Mei\ss ner St and D\"orfel B - D 1996
                 {\it J. Phys. A: Math. Gen.} {\bf 29} 1949
\bibitem{doerfel}D\"orfel B - D and Mei\ss ner St 1996
                 {\it J. Phys. A: Math. Gen.} {\bf 29} 6471
\bibitem{eckle}Woynarovich F and Eckle H-P 1987
                 {\it J. Phys. A: Math. Gen.} {\bf 20} L97
\bibitem{woy}Woynarovich F 1987
                 {\it Phys. Rev. Lett.} {\bf 59} 259, 1264 (erratum)
\bibitem{hamer}Hamer C J, Quispel G R W and Batchelor M T 1987
		 {\it J. Phys. A: Math. Gen.} {\bf 20} 5677
\bibitem{juettner}J\"uttner G and D\"orfel B - D 1993
                 {\it J. Phys. A: Math. Gen.} {\bf 26} 3105
\bibitem{yangyang}Yang C N and Yang C P 1966
		 {\it J. Phys. Rev.} {\bf 150} 327
\end{thebibliography}
\end{document}